# EXPANDING THE COMPUTATION OF MIXTURE MODELS BY THE USE OF HEREMITE POLYNOMIALS AND IDEALS

Andrew Clark

The Leeds Business School

University of Colorado at Boulder

Andrew.clark@colorado@edu

Abstract

Mixture models have found uses in many areas. To list a few: unsupervised learning, empirical Bayes, latent class and trait models. The current applications of mixture models to empirical data is limited to computing a mixture model from the same parametric family, e.g. Gaussians or Poissons. In this paper it is shown that by using Hermite polynomials and ideals, the modeling of a mixture process can be extended to include different families in terms of their cumulative distribution functions (cdfs)

Introduction

The current method of calculating a mixture model assumes the following:

> *N* random variables corresponding to observations, each assumed to be distributed according to a mixture of *K* components, with each component belonging to the same parametric family of distributions (e.g., all Gaussians, all Bernouillis) but with different parameters.

Therefore, when using current mixture modeling techniques, one is limited, analytically, to using just one family of distributions, such as a Gaussian, to compute a mixture model. There is a growing recognition amongst users of mixture models, such as civil engineers and risk managers, that the mixture models of interest to them are often combinations of two distinct families of distributions. For example, Bouchard and Potters[1] show that security returns are often a combination of Levy and Gaussian distributions while Wang et al[2] show the efficacy of using a mixture of t and Gaussian distributions in order to price the risk of credit default swaps and credit default obligations.

In order to effectively calculate a mixture model made up of 2 or more distinct families of distributions, we show that the combinations and related weights of distinct cdfs can be computed analytically via the use of Hermite polynomials and ideals. The methodology allows an arbitrarily number of distinct distributions to be tested simultaneously if no prior knowledge exists in terms of which distinct distribution families can or should be used.

---

[1] J-P Bouchard and M Potters, *Theory of Financial Risks*, Cambridge University Press, 2001, Chapter 2
[2] D. Wang, S.T. Rachev and F.J. Fabozzi, "Pricing Tranches of a CDO and a CDS Index: Recent Advances and Future Research", in *Risk Assessment*, G. Bol, S.T. Rachev, R. Wurth (eds.), Springer, 2009, pp. 263-286

Development

What follows is a fairly standard development in terms of algebraic geometry.

Given $\phi(x) = the\ cdf\ of\ x, where\ x\ are\ the\ observed\ data\ points, \exists\ a\ mixture\ model$
$s.t.\ \phi(x) = \sum_{j=1}^{n} \lambda_j \phi_j^*(x)$ where the $\lambda_j$ weights are assigned to their respective $\phi_j^*(x)$ cdfs which are being used to model $\phi(x)$. In addition, $\sum_i^j \lambda_j = 1$.

It is possible to compute the successive values of $\lambda_j\ given\ \phi_j^*(x)$ as long as $j < \infty$, i.e. there is a finite mixture model.

As Hermite polynomials exist for many known continuous and discrete distributions, such as the gamma family, we propose the use of such polynomials to solve for $\lambda_j$ in the equation $\phi(x) = \sum_{j=1}^{n} \lambda_j \phi_j^*(x)$ where $\phi_j^*(x)$ are Hermite polynomials.

Given that the distinct families of Hermites, $He_1, \ldots, H_j$, are $\in \mathbb{R}$, we need to compute the real roots, i.e. $\lambda_j$, the real variety, $V_\mathbb{R}(I)$, of the ideal $I = He_1, \ldots, H_j$. In this process, we will also find the real radical ideal, $\sqrt[\mathbb{R}]{I}$.

As known, $V_\mathbb{R}(I) = \{v \in \mathbb{R}^n\ \|\ f(y) = 0\ \forall\ f \in I\}$ and $\sqrt[\mathbb{R}]{I} = \{f \in \mathbb{R}(x)\ \|\ \exists n \in \mathbb{N} \wedge s_j \in \mathbb{R}(x) \wedge f^{2n} + \sum_j s_j^2 \in I\}$. Therefore, $I(V_\mathbb{R}(I)) = \{f \in \mathbb{R}(x)\ \|\ f(x) = 0\ \forall v \in V_\mathbb{R}(I)\}$ and $\sqrt[\mathbb{R}]{I} = I(V_\mathbb{R}(I))$.

One of the known ways of using of using *I* to compute the roots of polynomials is via eigenvalues. We will show the univariate case first.

Let $\lambda_j He_n = x^d - a_{d-1}x^{d-1} -, \ldots, -a_1 x - \wedge I$. Let $\beta = \{1, x, \ldots, x^{d-1}\}$ which is the linear basis of $\mathbb{R}(x)/I$.

Via the matrix $M_x$ of the multiplication (by x) operator in $\mathbb{R}(x)/I$, the eigenvalues of $M_x$ are the roots of $\lambda He_n$.

In the multivariate case, such as solving the $\lambda_j$'s for a finite mixture model of two or more $He_n$, assume $n < \infty$.

Let $m_f: \mathbb{R}(x)/I \to \mathbb{R}/I \wedge (p) \mapsto (f_p)$ are the multiplication by $f$ linear operator in $\mathbb{R}(x)/I$. Let $M_f$ be the matrix of $m_f$ in a base $\beta$ of $\mathbb{R}(x)/I$. The eigenvalues of $M_f = \{f(v) \| v \in V_\mathbb{C}(I)\}$. SDF (semidefinite programming) can be used to generate the roots. Therefore, determining $\lambda_j$'s of the Hermite polynomials can be solved using SDF.

Conclusions

In the brief development above, an analytical way of computing the weights of a mixture model, where the cdfs are from distinct families, is shown. As noted above, existing mixture methodologies are limited to each cdf belonging to the same parametric family of distributions. Using Hermite polynomials and computing ideals allows the investigator to mix distributions from distinct families.

A second potential use of the Hermite polynomial and ideal methodology would be investigative, i.e. part of an exploratory data analysis (EDA). Using the methodology above, two or more distinct cdfs could be used in a first pass EDA in order to determine what mix of cdfs would be of benefit. Such an analysis would be done prior to an investigation following the work of Wang et al, i.e. the investigator needs to determine what two cdfs, be they continuous and/or discrete, could be used to construct the mixture copula.

The methodology could also be used to confirm the results of any of the existing mixture model techniques, as long as the cdfs are not using the same number of moments. For example, a mixture of Gaussians which relies upon a model in which one cdf uses the first two moments and a second the first through fourth moments, can be computed and the weights compared.

An open question is the application of algebraic geometry to the calculation of copulas. Given that copulas are defined as a mixture of marginals, it may be possible to use the techniques of algebraic geometry to compute the weights of the marginals.